\documentclass[%
 reprint,
 amsmath,amssymb,
 aps,
]{revtex4-1}

\usepackage{graphicx}
\usepackage{dcolumn}
\usepackage{bm}
\usepackage{physics}
\usepackage[makeroom]{cancel}
\usepackage{siunitx}
\usepackage{multirow} 

\usepackage[dvipsnames]{xcolor}
\usepackage{soul}

\begin{document}

\preprint{APS/123-QED}

\title[]{Extrinsic Spin-Orbit Coupling and Spin Relaxation in Phosphorene}

\author{S. M. Farzaneh}
\email{farzaneh@nyu.edu}
\altaffiliation[]{}
\author{Shaloo Rakheja}
\affiliation{Department of Electrical and Computer Engineering, New York University, Brooklyn, NY 11201}

\begin{abstract}
An effective Hamiltonian is derived to describe the conduction band of monolayer black phosphorus (phosphorene) in the presence of spin-orbit coupling and external electric field.
Envelope function approximation along with symmetry arguments and Lowdin partitioning are utilized to derive extrinsic spin-orbit coupling. 
The resulting spin splitting appears in fourth order perturbation terms and is shown to be linear in both the magnitude of the external electric field and the strength of the atomic spin-orbit coupling, similar to the Bychkov-Rashba expression but with an in-plane anisotropy. 
The anisotropy depends on the coupling between conduction band and other bands both close and distant in energy. 
The spin relaxation of conduction electrons is then calculated within the Dyakonov-Perel mechanism where momentum scattering randomizes the polarization of a spin ensemble.
We show how the anisotropic Fermi contour and the anisotropic extrinsic spin splitting contribute to the anisotropy of spin-relaxation time.  
Scattering centers in the substrate are considered to be charged impurities with screened Coulomb potential. 
\end{abstract}
\maketitle

\section{introduction}
Extrinsic spin-orbit coupling induced by an external electric field in two-dimensional electron systems lifts the spin degeneracy while it preserves time-reversal symmetry. 
The induced spin splitting, which is proportional to the magnitude of the field and the crystal wavevector, enables control of spin through movement of charge and vice versa. 
This effect, which has enabled several phenomena and ideas in spintronics and beyond  \cite{manchon2015new}, was originally derived by Ohkawa and Uemura \cite{ohkawa1974quantized} for an inversion layer of zinc-blende crystals. Later, Vasko \cite{vasko1979spin}, and Bychkov and Rashba \cite{bychkov1984properties}, generalized the spin splitting for a two-dimensional electron system with an isotropic in-plane effective mass. 
Unlike Ohkawa and Uemura's derivation based on the Kane's model of zinc-blende crystals \cite{kane1957band}, the Vasko and Bychkov-Rashba spin splittings are phenomenological.
In this paper, we utilize envelope function approximation and symmetry arguments to derive the spin splitting for monolayer black phosphorus, which demonstrates a highly anisotropic in-plane effective mass and, therefore, makes the phenomenological description inapplicable. 

Black phosphorus, the most stable allotrope of phosphorus, is a layered material similar to graphite where van der Waals interaction binds individual layers together. 
Each monolayer, dubbed phosphorene, is a two-dimensional crystal with a puckered honeycomb structure which shares the symmetry properties of its bulk form denoted by the orthorhombic space group $Cmca$ \cite{hultgren1935atomic, brown1965refinement}.
A century after black phosphorus was discovered \cite{bridgman1914two}, phosphorene and its multilayer thin films were isolated \cite{li2014black, koenig2014electric, liu2014phosphorene, buscema2014fast, castellanos2014isolation, xia2014rediscovering} using mechanical exfoliation, which had been utilized earlier to isolate graphene \cite{novoselov2004electric}.
Similar to graphene, phosphorene consists of light atoms producing a spin-orbit coupling of $\sim$1 meV, which is weaker than that of conventional zinc-blende crystals. 
Hence, both graphene and phosphorene are expected to have a long spin-relaxation time $\sim$1 ns \cite{tombros2007electronic, avsar2017gate}, which could allow spin-polarized currents to flow macroscopic distances in these materials.
Unlike graphene, which is gapless, 
phosphorene is a semiconductor with a direct band gap of 1.73 eV \cite{li2017direct}, which enables a wide control over its carrier density.
Phosphorene also exhibits a large anisotropy in its band structure: the ratio of the in-plane effective mass of carriers along the armchair and zigzag directions is $\sim 0.1$ \cite{liu2014phosphorene}.
This band structure anisotropy is expected to result in anisotropic extrinsic spin-orbit coupling, also confirmed through first-principles calculations~\cite{popovic2015electronic, kurpas2016spin}.
It is shown \cite{ohkawa1974quantized} that in conventional semiconductors with a zinc-blende crystal the three upper valence bands, which are made of only $p$ orbitals at the band edge, induce the spin splitting in the conduction band made of $s$ orbital. 
However, in the case of phosphorene, it is not clear which bands couple to the conduction band to produce spin splitting. 
While previous works~\cite{popovic2015electronic, kurpas2016spin} on extrinsic spin-orbit coupling in phosphorene are solely based on first-principles calculations, this work focuses on analytic derivation of the spin-orbit splitting and demonstration of the bands involved from a group theoretical aspect. 
Utilizing the symmetry analysis of phosphorene, developed in Ref. \onlinecite{li2014electrons}, we derive an effective Hamiltonian for the extrinsic spin-orbit coupling in the conduction band by Lowdin partitioning and then quantify the contributions of different bands in the coefficients of spin-orbit coupling in zigzag and armchair directions.

Using the effective Hamiltonian we then study how the anisotropy impacts spin lifetime, a key measure of spin transport properties.
Spin-relaxation time characterizes the decay of the polarization of a non-equilibrium spin ensemble due to random fluctuations of a  magnetic field. 
Extrinsic spin-orbit coupling acts as an effective momentum-dependent magnetic field and causes the spins of conduction electrons undergoing momentum scattering to relax.
This mechanism, introduced by Dyakonov and Perel \cite{dyakonov1972spin}, has been used to develop closed-form solutions of the spin-relaxation time in isotropic semiconductors~\cite{fabian2007semiconductor, averkiev1999giant}.
However, when the effective mass of carriers is anisotropic, momentum scattering becomes anisotropic as well and must be accounted for numerically to determine spin-relaxation time accurately.
The anisotropic spin relaxation time has been calculated before \cite{kurpas2016spin} considering a constant momentum scattering. Here, we assume that the momentum scattering is anisotropic and, therefore, angle dependent.
Generalizing the Dyakonov-Perel mechanism in the case of phosphorene, we account for the anisotropy of momentum scattering and calculate the spin-relaxation time for spin ensembles with different initial polarization.
Our calculations assume that the temperature is much lower than the Fermi energy but much greater than the spin splitting i.e. $E_\text{F} \gg T \gg \Delta E_\text{SO}$. Therefore, only the electrons at the Fermi energy are taken into account and the spin-orbit coupling is treated as a perturbation.

\section{Extrinsic Spin-Orbit Coupling}
Considering the two-dimensional crystal of phosphorene with a periodic lattice potential $V_0(\vb*{r})$ lying on the $xy$ plane with the armchair edge along the $x$-direction and the zigzag edge along the $y$-direction, the Hamiltonian in the presence of Pauli spin-orbit coupling, $H_\text{SO}$, and a perpendicular electric field $V(z)=-e\mathcal{E}z$ is 
\begin{equation}
\label{eq:hamiltonian}
    \mathcal{H} = \underbrace{\frac{p^2}{2m_0} + V_0(\vb*{r})}_{H_0} + \underbrace{\frac{\hbar^2}{4m_0^2c^2}\vb*{p}\vdot\vb*{\sigma}\times\grad{V_0(\vb*{r})}}_{H_\text{SO}} + V(z)\cdot
\end{equation} 
Here, $\vb*{p}$ is the momentum operator, $\vb*{\sigma}$ is the vector of Pauli matrices, and $m_0$ is the mass of free electron .
We note that the contribution of $V(z)$ in $H_\text{SO}$ is neglected. 
Using envelope function approximation \cite{winkler2003spin} and Bloch's theorem in the $xy$ subspace, we can describe the solutions to the Schrodinger's equation as
\begin{equation}
\label{eq:wavefunction}
    \Psi_n(\vb*{r}) = e^{i\vb*{k}_\parallel\vdot\vb*{r}_\parallel} \sum_{l,\sigma}f_{nl\sigma}(z)u_{l\sigma\vb*{0}}(\vb*{r})\ket{\sigma},
\end{equation} 
where $\vb*{k}_{\parallel}=(k_x, k_y, 0)$ and  $\vb*{r}_{\parallel}=(x, y, 0)$ are the in-plane wavevector and position respectively, $f_{nl\sigma}(z)$ are the envelope functions and $u_{l\sigma\vb*{0}}(\vb*{r})\ket{\sigma}$ are the lattice-periodic Bloch functions at the band edge (i.e. $\Gamma$ point, $\vb*{k}_\parallel=0$) which provide a complete and orthonormal basis. 
Plugging Eq. \ref{eq:hamiltonian} and Eq. \ref{eq:wavefunction} into the Schrodinger's equation, multiplying with $\bra{\sigma'}u_{l'\sigma'\vb*{0}}^*(\vb*{r})$, and integrating over the primitive unit cell of the lattice, we arrive at the eigenvalue equation for the envelope functions, $Hf(z)=Ef(z)$,
\begin{equation}
\label{eq:eigenvalue}
\begin{split}
    & \sum_{l,\sigma} \bigg[\qty(E_{l}(\vb*{0}) + \frac{\hbar^2(k_\parallel^2 - \frac{d^2}{dz^2})}{2m_0} + V(z))\delta_{ll'}\delta_{\sigma\sigma'} + \Delta_{l'\sigma'l\sigma} \\
    & + \frac{\hbar}{m_0}(\vb*{k}_\parallel -  i\frac{d}{dz}\vu*{z})\vdot\vb*{P}_{l'\sigma'l\sigma}
    \bigg]f_{nl\sigma}(z) = E_nf_{nl'\sigma'}(z),
\end{split}
\end{equation}
where $E_{l}(\vb*{0})=\ev{H_0}{l}$ are the energies at the band edge and $\vb*{P}_{l'\sigma'l\sigma} = \mel{l'\sigma'}{\vb*{p} + \frac{\hbar}{4m_0c^2}\vb*{\sigma}\times\nabla{V_0}}{l\sigma} \approx \mel{l'}{\vb*{p}}{l}\delta_{\sigma\sigma'}$ are approximated by the matrix elements of momentum operator which couple different bands at the edge. 
This approximation is valid for light atoms such as phosphorus where the contribution of spin-orbit coupling is orders of magnitude smaller than that of momentum operator.
The matrix elements of $H_\text{SO}$ are denoted by $\Delta_{l'\sigma'l\sigma} = \mel{l'\sigma'}{H_\text{SO}}{l\sigma}$.
We rewrite the Pauli spin-orbit coupling $H_\text{SO}=(\xi\hbar/2)\vb*{\mathcal{L}}\vdot\vb*{\sigma}$, where $\vb*{\mathcal{L}}$ is the angular momentum operator, and the parameter $\xi$ represents the strength of atomic spin-orbit coupling and includes the average radial contribution of $\nabla V_0(\vb*{r})$.
Therefore, we rewrite $\Delta_{l'\sigma'l\sigma} = (\xi\hbar/2)\mel{l'\sigma'}{\vb*{\mathcal{L}}\vdot\vb*{\sigma}}{l\sigma}$.

The infinite-dimensional Hilbert space in Eq. \ref{eq:eigenvalue} can be reduced to a finite-dimensional one by considering only the bands that are in the vicinity of the Fermi energy. 
These bands are usually made up of orbitals of the valence electrons namely $s$ and $p$ atomic orbitals in phosphorus.  
Using atomic orbitals as a basis, Li \textit{et al.} \cite{li2014electrons} studied the symmetry properties of phosphorene and the orbital composition of the bands at the $\Gamma$ point. 
Following their study, we derive an effective Hamiltonian for the extrinsic spin-orbit coupling in the conduction band utilizing the theory of invariants and Lowdin partitioning.
There are eight irreducible representations (IRs) in the space group of phosphorene, $Cmca$, denoted by $\Gamma_i^{\pm}$ at the $\Gamma$ point of the Brillouin zone. 
Each band at the $\Gamma$ point can be labeled by one of the IRs of the space group.  
Table \ref{tb:character} in the appendix lists 
the characters of these IRs. 
The invariants corresponding to each IR is also listed in this table. 
According to the theory of invariants, the energy of each band should be an invariant of $\Gamma_1^+$. Therefore, various products of $k_x$, $k_y$, $\sigma_x$, and $\sigma_y$ could appear in the diagonalized effective Hamiltonian as long as the direct product of their corresponding IRs results in $\Gamma_1^+$. 
For instance, from Table \ref{tb:character} we can see that terms such as $k_zk_x\sigma_y$ and  $k_zk_y\sigma_x$ will be present in the effective Hamiltonian but terms such as $k_zk_x\sigma_x$ and  $k_zk_y\sigma_y$ will not. 
Since the external electric field $V(z)$ breaks the inversion symmetry in the $z$-direction, it does not commute with $k_z$ and, therefore, after replacing $k_z$ with $-id/dz$, their commutator $[k_z, V(z)]=-idV(z)/dz$ leads to the extrinsic spin-orbit coupling.

To derive the effective Hamiltonian for the conduction band $H_{cc}$, we first represent the Hamiltonian in Eq. \ref{eq:eigenvalue} as $H = H_0 + H'$ where $H_0$ contains band-edge and free-electron energies and $H'$ contains the electric field and the off-diagonal parts of $H$ namely k.p and spin-orbit terms.  
By applying Lowdin partitioning up to the fourth order perturbation, the coupling of conduction band with other bands is incorporated in the effective Hamiltonian. 
The $z$-dependence is averaged out afterwards to obtain the final two-dimensional effective Hamiltonian. 
The details of Lowdin partitioning and the derivation of the effective Hamiltonian is provided in Appendix \ref{ap:derivation}. 
The resulting effective Hamiltonian is 
\begin{equation}
\label{eq:effective}
    H_{cc} = \frac{\hbar^2k_x^2}{2m_x} + \frac{\hbar^2k_y^2}{2m_y} + \lambda_xk_x\sigma_y + \lambda_yk_y\sigma_x,
\end{equation}
where $m_x$ and $m_y$ are the in-plane effective masses written as
\begin{equation}
    \frac{1}{m_{x(y)}} = \frac{1}{m_0}\qty(1 + \frac{2}{m_0}\sum_{l}\frac{|P_{x(y),cl}|^2}{E_c - E_l}), 
\end{equation}
where the sum is over $l=\Gamma_2^+$ and $l=\Gamma_3^+$ for $x$ and $y$ directions, respectively. 
The coefficients of extrinsic spin-orbit coupling term, $\lambda_x$ and $\lambda_y$, are of the following form 
\begin{equation}
\label{eq:soc-coeff}
    \begin{split}
        \lambda_{x(y)} = & (\xi\hbar/2)\frac{dV(z)}{dz}\frac{\hbar^2}{m_0^2} \\ 
    \times & \sum_{l,l'}\Im{P_zL_{y(x)}P_{x(y)}}\frac{(E_c - E_l') \pm (E_c - E_l)}{(E_c - E_l)^2(E_c - E_l')^2},
    \end{split}
\end{equation}
where $P$ and $L$ are matrix elements of momentum and angular momentum operators that couple $c$, $l$, and $l'$ bands according to the invariants and their corresponding IRs. 
For instance, for $l=\Gamma_1^+$ and $l'=\Gamma_2^+$ ($c=\Gamma_4^-$ denotes the conduction band), the matrix elements appearing in $\lambda_x$ are $P_{z,cl}=\mel{c}{p_z}{l}$, $L_{y,ll'}=\mel{l}{\mathcal{L}_y}{l'}$, and  $P_{x,l'c}=\mel{l'}{p_x}{c}$. 
Equation \ref{eq:fourth} contains all possible terms that produce extrinsic spin-orbit coupling which is linear in spin-orbit strength $\xi$, electric field $dV(z)/dz$, and in-plane crystal momentum $k_{x(y)}$.  

\begin{figure}
        \includegraphics[width=2.7in]{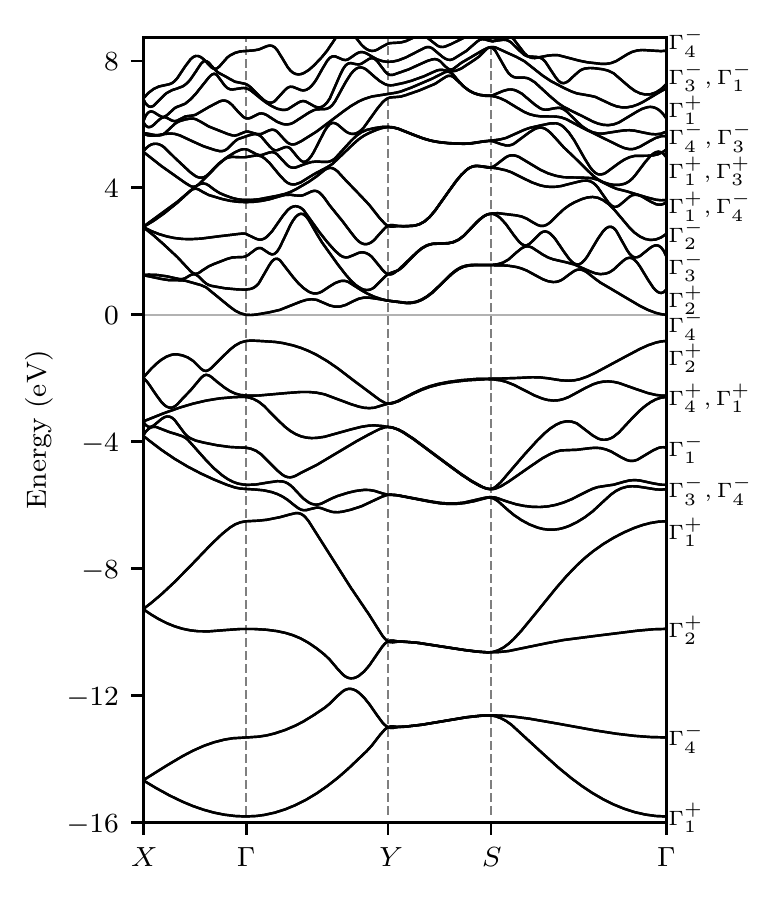}
    \vspace{-0.1in}
    \caption[example] 
    {Fully relativistic band structure of phosphorene. The lowest 24 bands are labeled at the $\Gamma$ point with respect to the irreducible representation of the space group of phosphorene.}
    \label{fig:band} 
\end{figure}

\begin{figure}
    \begin{tabular}{cc}
    \includegraphics[width=1.68in]{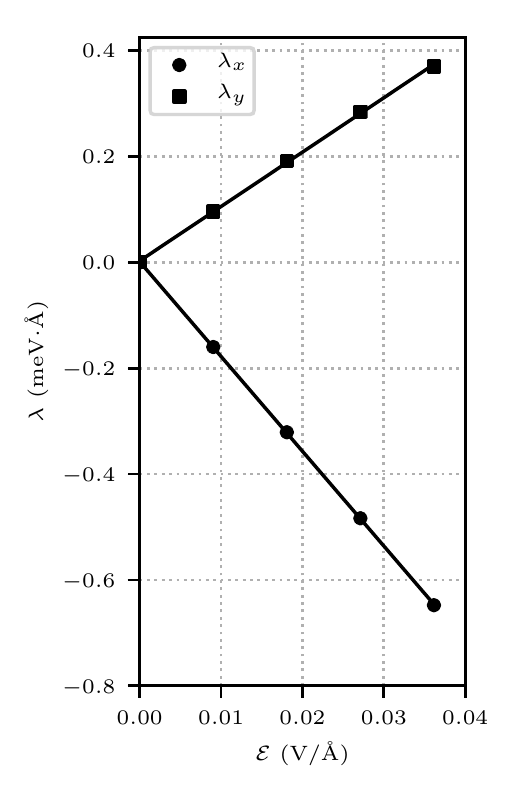} & \includegraphics[width=1.68in]{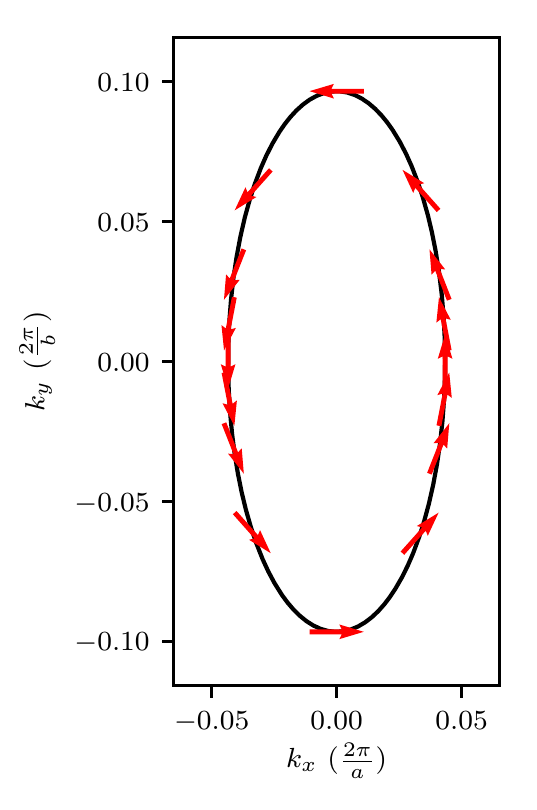}
    \end{tabular}
    \caption[example] 
    {(a) Coefficients of extrinsic spin splitting versus external electric field. For an electric field of $\mathcal{E}=0.1$ V/\AA, we obtain $\lambda_x=-1.79$ meV$\cdot$\AA and $\lambda_y=1.03$ meV$\cdot$\AA. (b) Expectation value of spin over the Fermi contour ($E_\text{f}=0.1$ eV) at $\mathcal{E}=0.36$ V/\AA.}
    \label{fig:efield} 
\end{figure}

First principle calculations within density functional theory (DFT) using projected augmented plane wave method implemented in Quantum ESPRESSO package \cite{QE-2017} were performed to 
verify the symmetry of the bands and also calculate the k.p parameters namely different $P$ and $L$ matrix elements appearing in Eq. \ref{eq:fourth}. 
Details of the DFT setup are provided in Table \ref{tb:dft}. 
Parameters of the crystal structure of phosphorene are listed in Table \ref{tb:structure}. 
Figure \ref{fig:band} illustrates the band structure of phosphorene for the lowest 24 bands for a path along the high symmetry points. The bands are labeled with their corresponding IRs which are consistent with previous calculations \cite{takao1981electronic, li2014electrons}.
Results show that $m_x=0.16m_0$ and $m_y=1.13m_0$ which is similar to what has been reported before \cite{popovic2015electronic, qiao2014high}.
Figure \ref{fig:efield}a plots $\lambda_x$ and $\lambda_y$ versus the external electric field. 
Quadratic terms with respect to the electric field appear only at high magnitudes, i.e. $\mathcal{E}>0.4$ V/\AA (not shown in the figure).
The expectation value of spin, over the Fermi contour, is illustrated in Fig. \ref{fig:efield}b.
As seen from the figure, the spin is not tangential to the Fermi contour in contrast to the isotropic case.
To verify Eq. \ref{eq:soc-coeff} produces the same anisotropy as that of Fig. \ref{fig:efield}a, the various $P$ and $L$ matrix elements for different bands were calculated. The contributions of different bands in the spin-orbit coefficients are listed in Table \ref{tb:coefficients}. 
The band number corresponding to each IR is mentioned as well to uniquely specify the first 24 diagonalized bands ordered from the lowest energy to the highest. For instance, the conduction band is labeled as $\Gamma_4^-$ and is the 11$^\text{th}$ band from the bottom in Fig. \ref{fig:band}. 
As seen from the table, only few of the bands contribute significantly to the spin-orbit coefficients. 
Also contributions from different bands possess different signs which cancel each other out. 
The anisotropy of spin-orbit coupling is determined by the $\lambda_x/\lambda_y$ ratio which requires the contributions from different bands to be summed up. As seen from Table \ref{tb:coefficients} this ratio is $\lambda_x/\lambda_y\approx -1.4$ which is close to the full first principle  result of $\lambda_x/\lambda_y\approx -1.7$ plotted in Fig. \ref{fig:efield}. 
The small discrepancy can be resolved as one adds other contributions from weakly coupled bands to the sum in Eq. \ref{eq:soc-coeff}.

\section{Spin Relaxation}
Utilizing the effective Hamiltonian in Eq. \ref{eq:effective}, we study the spin relaxation of conduction electrons. 
Decomposing the Hamiltonian $H_{cc} = H + H'$, we rewrite the extrinsic spin-orbit coupling term as $H'=\vb*{\Omega}_{\vb*{k}}\vdot\vb*{\sigma}$ where $\vb*{\Omega}_{\vb*{k}}=\lambda_yk_y\vu*{x} + \lambda_xk_x\vu*{y}$ is an effective magnetic field which is $\vb*{k}$-dependent. 
This effective magnetic field causes the electrons with different momenta to precess around different axes.
Therefore, scattering between different momenta randomizes the precession of a polarized spin ensemble and consequently leads to spin relaxation. 
This is the aforementioned Dyakonov-Perel mechanism. 

To calculate the spin-relaxation time, we follow a similar procedure as in Refs. \onlinecite{averkiev2002spin, fabian2007semiconductor}, but we specifically analyze an anisotropic Fermi contour with an anisotropic extrinsic spin-orbit coupling. 
Considering a polarized spin ensemble which is spatially-homogeneous and is described by a $\vb*{k}$-dependent density matrix $\rho_{\vb*{k}}$, the time evolution is given as the following kinetic equation~\cite{averkiev1999giant} 
\begin{equation}
    \label{eq:evolution}
    \pdv{\rho_{\vb*{k}}}{t} = -\frac{1}{i\hbar}[\rho_{\vb*{k}}, \vb*{\Omega}_{\vb*{k}}\vdot\vb*{\sigma}] -
    \sum_{\vb*{k'}\not={\vb* k}}W_{\vb*{kk'}}(\rho_{\vb*{k}} - \rho_{\vb*{ k'}})\,,
\end{equation}
where we used $[\rho_{\vb*{k}}, H] = 0$. Here $W_{\vb*{kk'}}$ is the probability density of transition between $\vb*{k}$ and $\vb*{k'}$ states. The first term on the right-hand side represents spin precession about $\vb*{\Omega}_{\vb*{k}}$, and the second term represents momentum scattering between incoming wavevector $\vb*{k}$ and outgoing wavevector $\vb*{k'}$. We assume that the density matrix can be decomposed as $\rho_{\vb*{k}}=\overline{\rho} + \rho'_{\vb*{ k}}$, where $\overline{\rho}$ is the average of density matrix over the Fermi contour, i.e. $\overline{\rho} = \ell^{-1}\int d\ell\rho_{\vb* k}$, where $\ell$ is the perimeter of the Fermi contour and $d\ell=d\theta\abs{\pdv*{\vb*{k}}{\theta}}$ is the differential arc length.
We assume that $\rho'_{\vb*{k}}$ is a small perturbation with zero average, i.e. $\overline{\rho}_{\vb*{k}}'=0$. 
Taking the average of Eq. \ref{eq:evolution} over the Fermi contour, we obtain
\begin{equation}
    \label{eq:evolution-1}
    \pdv{\overline{\rho}}{t} = 
    \frac{1}{i\hbar}\overline{[\rho'_{\vb*{k}}, \vb*{\Omega}_{\vb*{k}}\vdot\vb*{\sigma}]}\,,
\end{equation}
where we used the fact that $\overline{\vb*{\Omega}_{\vb*{k}}}$ is zero. The reason is that for each point $\vb*{k}$ on the Fermi contour, $-\vb*{k}$ is also on the Fermi contour. Since $\vb*{\Omega}_{\vb*{k}}$ is linear in $\vb*{k}$ and therefore an odd function of $\vb*{k}$, i.e. $\vb*{\Omega}_{-\vb*{k}}=-\vb*{\Omega}_{\vb*{k}}$, it averages to zero over the Fermi contour.
Applying the decomposition to Eq. \ref{eq:evolution} and dropping the terms containing product of $\vb*{\Omega}_{\vb*{k}}$ and $\rho'_{\vb*{k}}$, we can find the quasistatic value of $\rho'_{\vb*{k}}$, by setting $\pdv*{\rho'_{\vb*{k}}}{t}$ to zero, assuming that momentum relaxation is much faster than spin relaxation.
Therefore, 
\begin{equation}
    \label{eq:evolution-2}
    \frac{1}{i\hbar}[\overline{\rho}, \vb*{\Omega}_{\vb*{k}}\vdot\vb*{\sigma}] =
    \sum_{\vb*{k'}\not=\vb*{k}}W_{\vb*{kk'}}(\rho'_{\vb*{k}} - \rho'_{\vb* {k'}})\,\cdot
\end{equation}
Equations \ref{eq:evolution-1} and \ref{eq:evolution-2} are coupled and must be solved self-consistently. 
To do so, first we assume that the average spin polarization is in $\vu*{s}$ direction. 
Therefore, we can write $\overline{\rho} = (1 + \vu*{s}\vdot\vb*{\sigma})/2$.
It can be shown that $[\overline{\rho}, \vb*{\Omega}_{\vb*{k}}\vdot\vb*{\sigma}] = i(\vu*{ s}\times\vb*{\Omega}_{\vb*{k}})\vdot{\vb* \sigma}$.
Using Eq. \ref{eq:evolution-2}, we can solve for $\rho'_{\vb*{k}}$ iteratively using the following equation:
\begin{equation}
    \label{eq:rho_p}
    \rho'_{\vb*{k}} = \frac{\frac{1}{\hbar}(\vu*{s}\times\vb*{\Omega}_{\vb*{k}})\vdot{\vb* \sigma} + \sum_{\vb*{k'}\not=\vb*{k}}W_{\vb*{kk'}}\rho'_{\vb*{k'}}}{\sum_{\vb*{k'}\not=\vb*{k}}W_{\vb*{kk'}}}\,\cdot
\end{equation}
Plugging $\rho'_{\vb*{k}}$ into Eq. \ref{eq:evolution-1}, we calculate the rate of decay $\pdv* {\overline{\rho}}{t}$ or correspondingly $\dv*{\vu*{s}}{t}=-\vu*{s}/\tau_s$ which results in the spin-relaxation time $\tau_s$.

The collision sum in the continuum limit becomes an integral, i.e. $\sum_{\vb*{k'}\not=\vb*{k}}W_{\vb*{kk'}}\rightarrow A\int d^2\vb*{k'}(2\pi)^{-2}W_{\vb*{kk'}}$, where $A$ is the area of the system. 
Using Fermi's golden rule, the probability density of transition is given as $W_{\vb*{kk'}}=\frac{2\pi}{\hbar}N\abs{U_{\vb*{k}\vb*{k'}}}^2\delta(E({\vb* k}) - E({\vb*{k'}}))$, where $N$ is the number of scatterers and $U_{\vb*{k}\vb*{k'}}$ is the matrix element of the scattering potential. 
For long-range scattering potential varying slowly compared to the periodic lattice potential, $U_{\vb*{k}\vb*{k'}}=U(\vb*{k} - \vb*{k'})/A = U(\vb*{q})/A$, where $U(\vb*{q})$ is the Fourier transform of $U(\vb*{r})$.
In two-dimensional electron systems, the effective Coulomb potential in the Fourier domain is \cite{ando1982electronic}
\begin{equation}
    \label{eq:coulomb}
    U(\vb*{q}) = \frac{2\pi e^2}{\kappa(q + q_s)}e^{-qd},
\end{equation}
where $\kappa$ is the average relative permittivity, $d$ is the depth of the scattering center in the substrate, and $q_s\approx 2\sqrt{m_xm_y}e^2/\kappa \hbar^2$ is the Thomas-Fermi screening constant. 
The delta function in $W_{\vb*{kk'}}$ reduces the $k$-space integral to an integral over the Fermi contour. Therefore,
\begin{equation}
    \label{eq:momentum-relaxation-final}
    \sum_{\vb*{k'}\not=\vb*{k}}W_{\vb*{kk'}}\rightarrow
    \frac{n}{2\pi\hbar}
    \int d\ell'\frac{\abs{U\qty({\vb* q})}^2}{\abs{\grad E\qty(\vb*{k'})}}\,,
\end{equation}
where $n=N/A$ is the density of scatterers.  

\begin{figure}
    \begin{tabular}{cc}
        \includegraphics[width=1.68in]{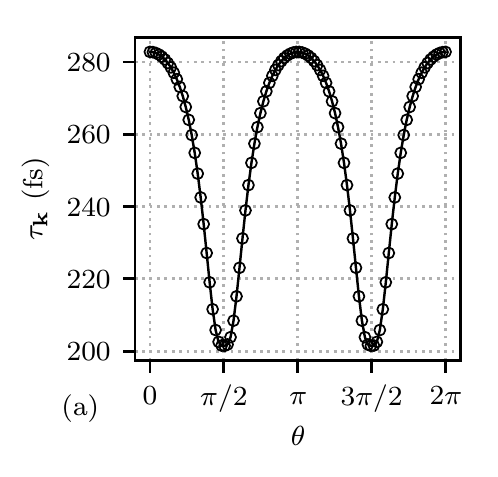} & 
        \includegraphics[width=1.68in]{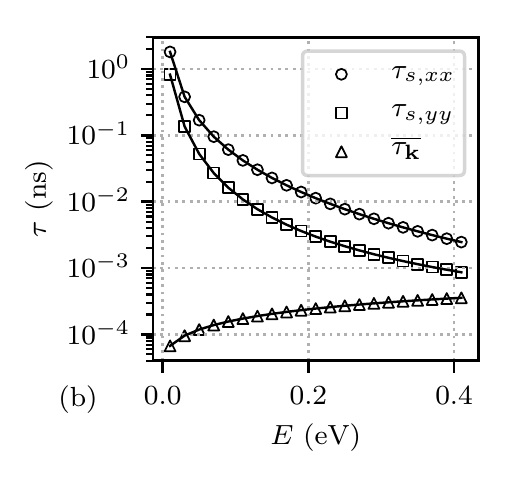}
    \end{tabular}
    \vspace{-0.1in}
    \caption[example] 
    {
    (a) $\vb*{k}$-dependent momentum scattering time for $n=10^{12}$cm$^{-2}$, $\kappa=2.45$, and $E=0.2$ eV. 
    (b) Spin-relaxation time along with the average momentum scattering time versus Fermi energy. The external electric field is assumed to be fixed at $\mathcal{E}=0.1$ V/\AA~ and the spin-orbit coefficients are $\lambda_x=-1.79$ meV$\cdot$\AA~ and $\lambda_y=1.03$ meV$\cdot$\AA. 
    }
    \label{fig:relax} 
\vspace{-0.1in}
\end{figure}

The $\vb*{k}$-dependent momentum scattering time, $\tau_{\vb*{k}}=1/\sum_{\vb*{k'}}W_{\vb*{kk'}}$, is depicted in Fig. \ref{fig:relax}a as a function of the polar angle for a typical value of charged impurity density \cite{yuan2015transport, liu2016mobility}, i.e. $n=10^{12}$ cm$^\text{-2}$. 
We assume that the monolayer is deposited on an SiO$_2$  substrate \cite{wang2015highly, li2014black} with relative permittivity of $\epsilon_\text{r}=3.9$.
As seen from the figure, the momentum scattering time shows a significant anisotropy which consequently affects the spin relaxation.

Figure \ref{fig:relax}b depicts the energy dependence of both average momentum scattering time, $\overline{\tau_{\vb*{k}}}$, and spin-relaxation time for two ensembles initially polarized in the x-direction, $\tau_{s,xx}$, and the y-directions, $\tau_{s,yy}$.
We assume that the external electric field is fixed at $\mathcal{E}=0.1$ V/\AA~ and the spin-orbit coefficients are $\lambda_x=-1.79$ meV$\cdot$\AA~and $\lambda_y=1.03$ meV$\cdot$\AA~ (from Fig. \ref{fig:efield}). 
Spin lifetime of the x-polarized ensemble is about 2-3 times longer than that of the y-polarized ensemble. This anisotropy increases with an increase in the Fermi energy.
The spin-relaxation time decreases with Fermi energy, while the average momentum scattering time increases with Fermi energy. This opposite energy dependence is a signature of the Dyakonov-Perel mechanism. 
We note that our assumptions in deriving Eqs. \ref{eq:evolution-1} and \ref{eq:evolution-2} are valid as long as the average momentum scattering occurs on a faster timescale compared to spin relaxation, i.e. for $E<0.4$ eV. 
We also note that $1/\tau_{s,xy}=1/\tau_{s,yx}=0$.
The spin relaxation for an ensemble polarized along the $\vu*{z}$ axis is always faster than in-plane directions (not shown in the figure). 
Replacing $\vu*{s}$ with $\vu*{z}$ in Eq. \ref{eq:rho_p}, we can see that $\rho'_{\vb*{k}}$ obtains both $\sigma_x$ and $\sigma_y$ components. Therefore, the corresponding spin relaxation rate is the sum of relaxation rates along the in-plane directions, i.e. $1/\tau_{s,zz} = 1/\tau_{s,xx} + 1/\tau_{s,yy}$.
We note that as the energy increases, the ratio of in-plane spin-relaxation times, $\tau_{s, yy}/\tau_{s, xx}$, increases as well.
The reason is that at higher energies screening becomes less effective as we can see from the screened Coulomb potential in Eq. \ref{eq:coulomb} which leads to a greater anisotropy in $\tau_{\vb*{k}}$. 
Finally, for an isotropic two-dimensional system, i.e. $m_x=m_y$ and $\lambda_x=\lambda_y$, we obtain $\tau_{s,xx}=\tau_{s,yy}=2\tau_{s,zz}$ which has been reported previously in the literature \cite{fabian2007semiconductor}.

\section{Summary}
Using envelope function approximation along with symmetry arguments and Lowdin partitioning an effective Hamiltonian was derived to describe the extrinsic spin-orbit coupling for the conduction electrons in phosphorene. 
Based on the theory of invariants, we determined the bands that are involved in generating extrinsic spin splitting. 
In contrast to the isotropic Bychkov-Rashba and Vasko spin splittings, phosphorene shows an anisotropic spin splitting which is characterized by two coefficients.
The coefficients are determined by the coupling between the conduction band and other bands. 
First-principles calculations were performed to verify the symmetry of the bands and also quantify the contribution of different bands in spin splitting of conduction electrons by calculating the parameters of the effective Hamiltonian. 
Given the effective Hamiltonian in the conduction band, we calculated the spin-relaxation time for a homogeneous polarized spin ensemble within a generalized Dyakonov-Perel mechanism. 
Our results show that spin-relaxation time is highly anisotropic in the plane of phosphorene. 
A spin ensemble polarized in the armchair direction (x-direction) relaxes about 2-3 times longer than a spin ensemble polarized in the zigzag direction (y-direction). 
The calculated spin lifetimes are comparable in magnitude to the recent experiment on phosphorene \cite{avsar2017gate} which is shown to be dominated by Elliott-Yafet mechanism. 
However, in order for this anisotropy to be detected experimentally, the Dyakonov-Perel mechanism needs to be dominant. 
Therefore, a significant electric field $\mathcal{E}\sim0.1$ V/\AA~ and a highly disordered sample with charged impurity density $n=10^{12}$ cm$^{-2}$ would be required.

\vspace{-0.1in}
\begin{acknowledgments}
The authors acknowledge the funding support from the MRSEC Program of the National Science Foundation under Award Number DMR-1420073.
\end{acknowledgments}

\appendix
\section{Lowdin Partitioning}
\label{ap:derivation}

\bgroup
\def\arraystretch{1.2}
\begin{table*}
\caption{Character table of the space group of phosphorene ($Cmca$: 64) for the $\Gamma$ point. The plus and minus signs denote the parity under spatial inversion $i$.}
\label{tb:character}
\begin{ruledtabular}
\begin{tabular}{ccccccccccc}
 & $\{E|0\}$ & $\{C_{2x}|\tau\}$ & $\{C_{2y}|0\}$ & $\{C_{2z}|\tau\}$ & $\{i|0\}$ & $\{R_{x}|\tau\}$ & $\{R_{y}|0\}$ & $\{R_{z}|\tau\}$ & Basis & Invariant \\
\hline 
$\Gamma_1^+$ & 1 & 1 & 1 & 1 & 1 & 1 & 1 & 1 & & $k_x^2 + k_y^2$\\
$\Gamma_2^+$ & 1 & -1 & 1 & -1 & 1 & -1 & 1 & -1 & $xz$ & $\sigma_y$\\
$\Gamma_3^+$ & 1 & 1 & -1 & -1 & 1 & 1 & -1 & -1 & $yz$ & $\sigma_x$\\
$\Gamma_4^+$ & 1 & -1 & -1 & 1 & 1 & -1 & -1 & 1 & $xy$ & $\sigma_z$\\
$\Gamma_1^-$ & 1 & 1 & 1 & 1 & -1 & -1 & -1 & -1 & & \\
$\Gamma_2^-$ & 1 & -1 & 1 & -1 & -1 & 1 & -1 & 1 & $y$ & $k_y$\\
$\Gamma_3^-$ & 1 & 1 & -1 & -1 & -1 & -1 & 1 & 1 & $x$ & $k_x$\\
$\Gamma_4^-$ & 1 & -1 & -1 & 1 & -1 & 1 & 1 & -1 & $z$ & $k_z=-id/dz$
\end{tabular}
\end{ruledtabular}
\end{table*}
\egroup

Lowdin partitioning \cite{lowdin1951note}, also known as quasi-degenerate perturbation theory, is a method for block diagonalizing a Hamiltonian with two sets of bands that weakly interact with each other. Here we are interested in the conduction band only and therefore calculate the effect of all other bands perturbatively as a separate set interacting with the conduction band. 
We assume that the Hamiltonian of the conduction band is denoted by $H_{cc}$ and the Hamiltonian in Eq. \ref{eq:eigenvalue} is decomposed as $H=H_0 + H'$ where 
\begin{equation}
\begin{split}
    H_{0,ll'} = (E_l + \frac{\hbar^2(k_\parallel^2 - (d/dz)^2}{2m_0})\delta_{ll'},\\
    H_{ll'}' = V(z)\delta_{ll'} + \Delta_{ll'} + \frac{\hbar}{m_0}(\vb*{k}_\parallel - i\frac{d}{dz})\cdot\vb*{P}_{ll's}\cdot
\end{split}
\end{equation}
Based on Lowdin partitioning, the leading four terms that comprise $H_{cc}$ which are given as follows\cite{winkler2003spin}. 
\begin{subequations}
    \begin{flalign}
        & H_{cc}^{(0)} = H_{0,cc},&
    \end{flalign}
    \begin{flalign}
        & H_{cc}^{(1)} = H_{cc}',& 
    \end{flalign}
    \begin{flalign}
        & H_{cc}^{(2)} = \sum_{l}\frac{H'_{cl}H'_{lc}}{E_c - E_l},& 
    \end{flalign}
    \begin{flalign}
        H_{cc}^{(3)} =   -\frac{1}{2} & \sum_{l}\frac{H'_{cc}H'_{cl}H'_{lc}}{(E_c - E_l)^2} -\frac{1}{2}\sum_{l}\frac{H'_{cl}H'_{lc}H'_{cc}}{(E_c - E_l)^2}&& \\\nonumber
        + &\sum_{l,l'}\frac{H'_{cl}H'_{ll'}H'_{l'c}}{(E_c - E_l)(E_c - E_{l'})}
    \end{flalign}
    \begin{flalign}
        H_{cc}^{(4)} = \frac{1}{2} &\sum_{l}\frac{H'_{cc}H'_{cc}H'_{cl}H'_{lc} + H'_{cl}H'_{lc}H'_{cc}H'_{cc}}{(E_c - E_l)^3} && \\\nonumber
        -\frac{1}{2} &\sum_{l,l'}\qty(H'_{cl}H'_{ll'}H'_{l'c}H'_{cc} + H'_{cc}H'_{cl}H'_{ll'}H'_{l'c}) && \\ \nonumber
        & \times \qty(\frac{(E_{c} - E_{l'}) + (E_c - E_{l})}{(E_{c} - E_l)^2(E_{c} - E_{l'})^2}) && \\ \nonumber
        -\frac{1}{2} &\sum_{l,l'}H'_{cl}H'_{lc}H'_{cl'}H'_{l'c}\qty(\frac{(E_{c} - E_{l'}) + (E_c - E_{l})}{(E_{c} - E_l)^2(E_{c} - E_{l'})^2}) && \\ \nonumber
        + &\sum_{l,l',l''}\frac{H'_{cl}H'_{ll'}H'_{l'l''}H'_{l''c}}{(E_c - E_{l})(E_{c} - E_{l'})(E_{c} - E_{l''})}
    \end{flalign}
\end{subequations}
The zeroth order term is similar to the energy dispersion of a free electron confined in two-dimensions i.e. $H_{cc}^{(0)} = H_{cc}^0 = E_c + \hbar^2(k_\parallel^2 - (d/dz)^2)/2m_0$. 
The first order term is similarly given by $H'$ as $H_{cc}^{(1)} = V(z)$ which vanishes upon taking the expectation value over the $z$ direction. 
The expectation value is $\langle H_{cc}^{(1)}\rangle=\int dzf^*_c(z)V(z)f_c(z)=0$ which is a direct consequence of inversion symmetry of the unperturbed envelope function $f_c(z)$. More generally, higher order terms that contain odd powers of $V(z)$ or $k_z=-id/dz$ vanish upon averaging over $z$. This makes the calculation of the higher order terms more convenient as many terms become zero. 
The second order term contains the effective mass terms, subband energy in the $z$ direction, and other terms quadratic in spin-orbit strength which we discard. 
\begin{equation}
\begin{split}
    H_{cc}^{(2)} = \frac{\hbar^2}{m_0^2}\Bigg[k_x^2 &\sum_{l=\Gamma_2^+}\frac{|P_{x,cl}|^2}{E_c - E_l} 
    + k_y^2\sum_{l=\Gamma_3^+}\frac{|P_{y,cl}|^2}{E_c - E_{l}} \\ 
    - \frac{d^2}{dz^2} &\sum_{l=\Gamma_1^+}\frac{|P_{z,cl}|^2}{E_c - E_l}\Bigg] + \mathcal{O}(\xi^2)\sigma_0\cdot
\end{split}    
\end{equation}
The first sum in the third order term vanishes upon averaging over $z$ for it contains $H'_{cc}=V(z)$ in all of its terms. The same is true for the second sum although it is more subtle as the terms with $l=\Gamma_1^+$ are operators affecting $H'_{cc}=V(z)$. In other words, since $c\otimes l = \Gamma_4^-\otimes\Gamma_1^+=\Gamma_4^-$, $H_{cl}$ contains $k_z$ invariant, i.e.  $H_{cl}=(\hbar/m_0)(-id/dz)P_{z,cl}$. Therefore, 
\begin{equation}
\begin{split}
    H'_{cl}H'_{lc}H'_{cc} = - \frac{\hbar^2|P_{z,cl}|^2}{m_0^2}\Bigg(\frac{d^2V(z)}{dz^2} 
    + 2\frac{dV(z)}{dz}\frac{d}{dz} \\+ V(z)\frac{d^2}{dz^2}\Bigg)\cdot
\end{split}
 \end{equation}
All the terms on the right hand side vanish upon averaging out $z$. 
Similarly, the third sum contains only odd terms in $V(z)$ and $d/dz$ which vanish as well. It also produces some high order terms in momentum and spin-orbit strength, i.e. $\langle H_{cc}^{(3)}\rangle=\mathcal{O}(\xi k_xk_y)\sigma_z + \mathcal{O}(\xi^3)\sigma_0$, which are negligible. 
The extrinsic spin orbit coupling, which couples electric field, in-plane momentum, and spin-orbit together, emerges from fourth order terms. 
The first sum in the fourth order term produces terms that are quadratic in electric field $(dV(z)/dz)^2$ which are negligible and appear only at high electric fields. 
The third sum produces other negligible high order terms i.e.  $\mathcal{O}(k_x^4) + \mathcal{O}(k_y^4) + \mathcal{O}(k_x^2d^2/dz^2) + \mathcal{O}(k_y^2d^2/dz^2) + \mathcal{O}(d^4/dz^4) + \mathcal{O}(\xi^2k_x^2) + \mathcal{O}(\xi^2k_y^2)+ \mathcal{O}(\xi^2d^2/dz^2) + \mathcal{O}(\xi^4)$. 
In the second sum, only terms containing both $V(z)$ and $d/dz$ survive. Other terms are either quadratic in $V(z)$ or zero after averaging out $z$. 
First order terms in $V(z)$ can be found in the fourth sum where either $l=l'$ or $l'=l''$. Other terms stemming from the fourth sum are negligible for they contain higher orders of electric field, spin-orbit, or momentum. 
Therefore, we can approximate the fourth sum as 
\begin{equation}
    \approx \sum_{l,l''}\frac{H'_{cl}H'_{ll}H'_{ll''}H'_{l''c}}{(E_c - E_{l})^2(E_{c} - E_{l''})} + 
    \sum_{l,l'}\frac{H'_{cl}H'_{ll'}H'_{l'l'}H'_{l'c}}{(E_c - E_{l})(E_{c} - E_{l'})^2}\cdot
\end{equation}
Since $H_{cc}=H_{ll}=V(z)$ for all $l$, we can combine the second and the fourth sum together. It might seem that they would cancel each other completely. However, since $V(z)$ and $d/dz$ do not commute with each other, their commutator, i.e. $[d/dz, V(z)]=dV(z)/dz$, survives the summation. We can see that this is the direct consequence of the broken inversion symmetry by $V(z)$ which leads to the extrinsic spin-orbit term. 
Finally, based on the multiplication of group elements we obtain the non-zero terms that couple electric field, momentum in the $z$ direction, an in-plane momentum, and the angular momentum perpendicular to the rest.
Therefore, we obtain the effective Hamiltonian for extrinsic spin-orbit coupling averaged over $z$ direction. 
\begin{equation}
\label{eq:fourth}
\begin{split}
    \ev{H_{cc}^{(4)}} & \approx (\xi\hbar/2) \frac{dV(z)}{dz}\frac{\hbar^2}{m_0^2}\Bigg[ \\  
    k_x\sigma_y & \sum_{\substack{l=\Gamma_1^+\\l'=\Gamma_2^+}}\Im{P_{z,cl}L_{y,ll'}P_{x,l'c}}\frac{(E_{c} - E_{l'}) + (E_c - E_{l})}{(E_c - E_{l})^2(E_{c} - E_{l'})^2}\\
    + k_x\sigma_y & \sum_{\substack{l=\Gamma_1^+\\l'=\Gamma_3^-}}\Im{P_{z,cl}P_{x,ll'}L_{y,l'c}}\frac{(E_{c} - E_{l'}) + (E_c - E_{l})}{(E_c - E_{l})^2(E_{c} - E_{l'})^2}\\
    - k_x\sigma_y & \sum_{\substack{l=\Gamma_2^+\\l'=\Gamma_3^-}}\Im{P_{x,cl}P_{z,ll'}L_{y,l'c}}\frac{(E_{c} - E_{l'}) - (E_c - E_{l})}{(E_c - E_{l})^2(E_{c} - E_{l'})^2}\\
    + k_y\sigma_x & \sum_{\substack{l=\Gamma_1^+\\l'=\Gamma_3^+}}\Im{P_{z,cl}L_{x,ll'}P_{y,l'c}}\frac{(E_{c} - E_{l'}) + (E_c - E_{l})}{(E_c - E_{l})^2(E_{c} - E_{l'})^2} \\
    + k_y\sigma_x & \sum_{\substack{l=\Gamma_1^+\\l'=\Gamma_2^-}}\Im{P_{z,cl}P_{y,ll'}L_{x,l'c}}\frac{(E_{c} - E_{l'}) + (E_c - E_{l})}{(E_c - E_{l})^2(E_{c} - E_{l'})^2} \\
    - k_y\sigma_x & \sum_{\substack{l=\Gamma_3^+\\l'=\Gamma_2^-}}\Im{P_{y,cl}P_{z,ll'}L_{x,l'c}}\frac{(E_{c} - E_{l'}) - (E_c - E_{l})}{(E_c - E_{l})^2(E_{c} - E_{l'})^2}\Bigg]\\
    & = (\xi\hbar/2)\frac{dV(z)}{dz}\frac{\hbar^2}{m_0^2}\sum_{l,l'} \alpha_{x,ll'} k_x\sigma_y + \alpha_{y,ll'} k_y\sigma_x\cdot
\end{split}
\end{equation}
As expected, this term is linear in spin-orbit strength $\xi$, electric field $dV(z)/dz$, and in-plane momentum $k_x$, $k_y$.  
Using first principle calculations we quantify the terms in Eq. \ref{eq:fourth} by calculating the matrix elements of momentum and angular momentum operator. 
The importance of each term is determined by the strength of coupling between the bands involved and also the energy difference between them. 
Table \ref{tb:coefficients} lists all significant nonzero terms and their corresponding coefficients.
The sum over all contributions (last row) determines the anisotropy  $\sum\alpha_{y,ll'}/\sum\alpha_{x,ll'}\approx -1.4~$.

\begin{table}[h]
\caption{First principle calculations setup}
\label{tb:dft}
\begin{ruledtabular}
\begin{tabular}{p{1.7in}p{1.6in}}
Pseudopotential Type & Ultrasoft, Fully Relativistic \\
Exchange-Correlation Function & PBE \\
Kinetic Energy Cutoff & 34.0 Ry\\
Charge Density Cutoff & 136.0 Ry\\
Convergence Threshold & $10^{-6}$ Ry\\
$k$-Point Grid & Monkhrost $12\times12\times1$ \\
\# of bands & 24 \\ 
Interlayer Spacing & $31.22=$ \AA 
\end{tabular}
\end{ruledtabular}
\end{table}
t
\def\arraystretch{1.2}
\begin{table}[h]
\caption{Parameters of the crystal structure of bulk and monolayer black phosphorus based on the definition in Ref. \onlinecite{takao1981electronic}.}
\label{tb:structure}
\begin{ruledtabular}
\begin{tabular}{p{0.5in}cccccc}
& $a$ {\footnotesize(\AA)} & $b$ {\footnotesize(\AA)} & $d_1$ {\footnotesize(\AA)} & $d_2$ {\footnotesize(\AA)} & $\alpha_1$ & $\alpha_2$\\
\hline
Bulk & 4.376 & 3.314 & 2.224 & 2.244 & 96.34$^\circ$ & 102.09$^\circ$ \\
Monolayer & 4.552 & 3.306 & 2.224 & 2.262 & 93.89$^\circ$ & 102.98$^\circ$ \\
\end{tabular}
\end{ruledtabular}
\end{table}

\begin{table}[h]    
\caption{Contributions of different bands in the extrinsic spin-orbit coupling quantified by coefficients $\alpha_{x,ll'}$ and $\alpha_{y,ll'}$ according to Eq. \ref{eq:fourth}. The values are obtained from first principle calculations and are in atomic units. The last row lists the total contributions of the bands which determines the anisotropy between extrinsic spin-orbit coefficients.}
\label{tb:coefficients}
\begin{ruledtabular}
\begin{tabular}{ccccrr}
    $l$ & band\# & $l'$ & band\# & $\alpha_{x,ll'}$ & $\alpha_{y,ll'}$\\
    \hline
    \multirow{5}{*}{$\Gamma_1^+$} & 4 & \multirow{5}{*}{$\Gamma_2^+$} & 10 & -71.24 & \\
     & 4 & & 12 & 2.95 & \\
     & 9 & & 10 & 10.08 & \\
     & 15 & & 10 & -6.39 & \\
     & 17 & & 10 & -5.68 & \\
     \hline
     \multirow{2}{*}{$\Gamma_1^+$} & 17 & \multirow{2}{*}{$\Gamma_3^-$} & \multirow{2}{*}{13} & -2.58 & \\
     & 21 & & & -5.75 & \\
     \hline
     \multirow{4}{*}{$\Gamma_2^+$} & \multirow{4}{*}{10} & \multirow{4}{*}{$\Gamma_3^-$} & 5 & 32.84 & \\
     & & & 13 & -2.76 & \\
     & & & 20 & 46.36 & \\
     & & & 22 & -8.77 & \\
     \hline 
     $\Gamma_1^+$ & & $\Gamma_3^+$ & & & $<$ 1\\
     \hline 
     \multirow{3}{*}{$\Gamma_1^+$} & 4 & \multirow{3}{*}{$\Gamma_2^-$} & \multirow{3}{*}{14} & & 2.10\\
     & 17 & & & & -2.76 \\
     & 21 & & & & -7.02\\ 
     \hline 
     $\Gamma_3^+$ & & $\Gamma_2^-$ & & & $<$ 1\\
     \hline 
     \multicolumn{4}{c}{$\sum_{l,l'}$} & -10.95 & 7.67
\end{tabular}
\end{ruledtabular}
\end{table}    

\clearpage
\bibliography{main}

\begin{thebibliography}{34}%
\makeatletter
\providecommand \@ifxundefined [1]{%
 \@ifx{#1\undefined}
}%
\providecommand \@ifnum [1]{%
 \ifnum #1\expandafter \@firstoftwo
 \else \expandafter \@secondoftwo
 \fi
}%
\providecommand \@ifx [1]{%
 \ifx #1\expandafter \@firstoftwo
 \else \expandafter \@secondoftwo
 \fi
}%
\providecommand \natexlab [1]{#1}%
\providecommand \enquote  [1]{``#1''}%
\providecommand \bibnamefont  [1]{#1}%
\providecommand \bibfnamefont [1]{#1}%
\providecommand \citenamefont [1]{#1}%
\providecommand \href@noop [0]{\@secondoftwo}%
\providecommand \href [0]{\begingroup \@sanitize@url \@href}%
\providecommand \@href[1]{\@@startlink{#1}\@@href}%
\providecommand \@@href[1]{\endgroup#1\@@endlink}%
\providecommand \@sanitize@url [0]{\catcode `\\12\catcode `\$12\catcode
  `\&12\catcode `\#12\catcode `\^12\catcode `\_12\catcode `\%12\relax}%
\providecommand \@@startlink[1]{}%
\providecommand \@@endlink[0]{}%
\providecommand \url  [0]{\begingroup\@sanitize@url \@url }%
\providecommand \@url [1]{\endgroup\@href {#1}{\urlprefix }}%
\providecommand \urlprefix  [0]{URL }%
\providecommand \Eprint [0]{\href }%
\providecommand \doibase [0]{http://dx.doi.org/}%
\providecommand \selectlanguage [0]{\@gobble}%
\providecommand \bibinfo  [0]{\@secondoftwo}%
\providecommand \bibfield  [0]{\@secondoftwo}%
\providecommand \translation [1]{[#1]}%
\providecommand \BibitemOpen [0]{}%
\providecommand \bibitemStop [0]{}%
\providecommand \bibitemNoStop [0]{.\EOS\space}%
\providecommand \EOS [0]{\spacefactor3000\relax}%
\providecommand \BibitemShut  [1]{\csname bibitem#1\endcsname}%
\let\auto@bib@innerbib\@empty
\bibitem [{\citenamefont {Manchon}\ \emph {et~al.}(2015)\citenamefont
  {Manchon}, \citenamefont {Koo}, \citenamefont {Nitta}, \citenamefont
  {Frolov},\ and\ \citenamefont {Duine}}]{manchon2015new}%
  \BibitemOpen
  \bibfield  {author} {\bibinfo {author} {\bibfnamefont {A.}~\bibnamefont
  {Manchon}}, \bibinfo {author} {\bibfnamefont {H.~C.}\ \bibnamefont {Koo}},
  \bibinfo {author} {\bibfnamefont {J.}~\bibnamefont {Nitta}}, \bibinfo
  {author} {\bibfnamefont {S.}~\bibnamefont {Frolov}}, \ and\ \bibinfo {author}
  {\bibfnamefont {R.}~\bibnamefont {Duine}},\ }\href@noop {} {\bibfield
  {journal} {\bibinfo  {journal} {Nature materials}\ }\textbf {\bibinfo
  {volume} {14}},\ \bibinfo {pages} {871} (\bibinfo {year} {2015})}\BibitemShut
  {NoStop}%
\bibitem [{\citenamefont {Ohkawa}\ and\ \citenamefont
  {Uemura}(1974)}]{ohkawa1974quantized}%
  \BibitemOpen
  \bibfield  {author} {\bibinfo {author} {\bibfnamefont {F.~J.}\ \bibnamefont
  {Ohkawa}}\ and\ \bibinfo {author} {\bibfnamefont {Y.}~\bibnamefont
  {Uemura}},\ }\href@noop {} {\bibfield  {journal} {\bibinfo  {journal}
  {Journal of the Physical Society of Japan}\ }\textbf {\bibinfo {volume}
  {37}},\ \bibinfo {pages} {1325} (\bibinfo {year} {1974})}\BibitemShut
  {NoStop}%
\bibitem [{\citenamefont {Vas’ko}(1979)}]{vasko1979spin}%
  \BibitemOpen
  \bibfield  {author} {\bibinfo {author} {\bibfnamefont {F.}~\bibnamefont
  {Vas’ko}},\ }\href@noop {} {\bibfield  {journal} {\bibinfo  {journal} {JETP
  Lett}\ }\textbf {\bibinfo {volume} {30}},\ \bibinfo {pages} {541} (\bibinfo
  {year} {1979})}\BibitemShut {NoStop}%
\bibitem [{\citenamefont {Bychkov}\ and\ \citenamefont
  {Rashba}(1984)}]{bychkov1984properties}%
  \BibitemOpen
  \bibfield  {author} {\bibinfo {author} {\bibfnamefont {Y.~A.}\ \bibnamefont
  {Bychkov}}\ and\ \bibinfo {author} {\bibfnamefont {E.}~\bibnamefont
  {Rashba}},\ }\href@noop {} {\bibfield  {journal} {\bibinfo  {journal} {JETP
  lett}\ }\textbf {\bibinfo {volume} {39}},\ \bibinfo {pages} {78} (\bibinfo
  {year} {1984})}\BibitemShut {NoStop}%
\bibitem [{\citenamefont {Kane}(1957)}]{kane1957band}%
  \BibitemOpen
  \bibfield  {author} {\bibinfo {author} {\bibfnamefont {E.~O.}\ \bibnamefont
  {Kane}},\ }\href@noop {} {\bibfield  {journal} {\bibinfo  {journal} {Journal
  of Physics and Chemistry of Solids}\ }\textbf {\bibinfo {volume} {1}},\
  \bibinfo {pages} {249} (\bibinfo {year} {1957})}\BibitemShut {NoStop}%
\bibitem [{\citenamefont {Hultgren}\ \emph {et~al.}(1935)\citenamefont
  {Hultgren}, \citenamefont {Gingrich},\ and\ \citenamefont
  {Warren}}]{hultgren1935atomic}%
  \BibitemOpen
  \bibfield  {author} {\bibinfo {author} {\bibfnamefont {R.}~\bibnamefont
  {Hultgren}}, \bibinfo {author} {\bibfnamefont {N.}~\bibnamefont {Gingrich}},
  \ and\ \bibinfo {author} {\bibfnamefont {B.}~\bibnamefont {Warren}},\
  }\href@noop {} {\bibfield  {journal} {\bibinfo  {journal} {The Journal of
  Chemical Physics}\ }\textbf {\bibinfo {volume} {3}},\ \bibinfo {pages} {351}
  (\bibinfo {year} {1935})}\BibitemShut {NoStop}%
\bibitem [{\citenamefont {Brown}\ and\ \citenamefont
  {Rundqvist}(1965)}]{brown1965refinement}%
  \BibitemOpen
  \bibfield  {author} {\bibinfo {author} {\bibfnamefont {A.}~\bibnamefont
  {Brown}}\ and\ \bibinfo {author} {\bibfnamefont {S.}~\bibnamefont
  {Rundqvist}},\ }\href@noop {} {\bibfield  {journal} {\bibinfo  {journal}
  {Acta Crystallographica}\ }\textbf {\bibinfo {volume} {19}},\ \bibinfo
  {pages} {684} (\bibinfo {year} {1965})}\BibitemShut {NoStop}%
\bibitem [{\citenamefont {Bridgman}(1914)}]{bridgman1914two}%
  \BibitemOpen
  \bibfield  {author} {\bibinfo {author} {\bibfnamefont {P.}~\bibnamefont
  {Bridgman}},\ }\href@noop {} {\bibfield  {journal} {\bibinfo  {journal}
  {Journal of the American Chemical Society}\ }\textbf {\bibinfo {volume}
  {36}},\ \bibinfo {pages} {1344} (\bibinfo {year} {1914})}\BibitemShut
  {NoStop}%
\bibitem [{\citenamefont {Li}\ \emph {et~al.}(2014)\citenamefont {Li},
  \citenamefont {Yu}, \citenamefont {Ye}, \citenamefont {Ge}, \citenamefont
  {Ou}, \citenamefont {Wu}, \citenamefont {Feng}, \citenamefont {Chen},\ and\
  \citenamefont {Zhang}}]{li2014black}%
  \BibitemOpen
  \bibfield  {author} {\bibinfo {author} {\bibfnamefont {L.}~\bibnamefont
  {Li}}, \bibinfo {author} {\bibfnamefont {Y.}~\bibnamefont {Yu}}, \bibinfo
  {author} {\bibfnamefont {G.~J.}\ \bibnamefont {Ye}}, \bibinfo {author}
  {\bibfnamefont {Q.}~\bibnamefont {Ge}}, \bibinfo {author} {\bibfnamefont
  {X.}~\bibnamefont {Ou}}, \bibinfo {author} {\bibfnamefont {H.}~\bibnamefont
  {Wu}}, \bibinfo {author} {\bibfnamefont {D.}~\bibnamefont {Feng}}, \bibinfo
  {author} {\bibfnamefont {X.~H.}\ \bibnamefont {Chen}}, \ and\ \bibinfo
  {author} {\bibfnamefont {Y.}~\bibnamefont {Zhang}},\ }\href@noop {}
  {\bibfield  {journal} {\bibinfo  {journal} {Nature nanotechnology}\ }\textbf
  {\bibinfo {volume} {9}},\ \bibinfo {pages} {372} (\bibinfo {year}
  {2014})}\BibitemShut {NoStop}%
\bibitem [{\citenamefont {Koenig}\ \emph {et~al.}(2014)\citenamefont {Koenig},
  \citenamefont {Doganov}, \citenamefont {Schmidt}, \citenamefont
  {Castro~Neto},\ and\ \citenamefont {Oezyilmaz}}]{koenig2014electric}%
  \BibitemOpen
  \bibfield  {author} {\bibinfo {author} {\bibfnamefont {S.~P.}\ \bibnamefont
  {Koenig}}, \bibinfo {author} {\bibfnamefont {R.~A.}\ \bibnamefont {Doganov}},
  \bibinfo {author} {\bibfnamefont {H.}~\bibnamefont {Schmidt}}, \bibinfo
  {author} {\bibfnamefont {A.}~\bibnamefont {Castro~Neto}}, \ and\ \bibinfo
  {author} {\bibfnamefont {B.}~\bibnamefont {Oezyilmaz}},\ }\href@noop {}
  {\bibfield  {journal} {\bibinfo  {journal} {Applied Physics Letters}\
  }\textbf {\bibinfo {volume} {104}},\ \bibinfo {pages} {103106} (\bibinfo
  {year} {2014})}\BibitemShut {NoStop}%
\bibitem [{\citenamefont {Liu}\ \emph {et~al.}(2014)\citenamefont {Liu},
  \citenamefont {Neal}, \citenamefont {Zhu}, \citenamefont {Luo}, \citenamefont
  {Xu}, \citenamefont {Tom{\'a}nek},\ and\ \citenamefont
  {Ye}}]{liu2014phosphorene}%
  \BibitemOpen
  \bibfield  {author} {\bibinfo {author} {\bibfnamefont {H.}~\bibnamefont
  {Liu}}, \bibinfo {author} {\bibfnamefont {A.~T.}\ \bibnamefont {Neal}},
  \bibinfo {author} {\bibfnamefont {Z.}~\bibnamefont {Zhu}}, \bibinfo {author}
  {\bibfnamefont {Z.}~\bibnamefont {Luo}}, \bibinfo {author} {\bibfnamefont
  {X.}~\bibnamefont {Xu}}, \bibinfo {author} {\bibfnamefont {D.}~\bibnamefont
  {Tom{\'a}nek}}, \ and\ \bibinfo {author} {\bibfnamefont {P.~D.}\ \bibnamefont
  {Ye}},\ }\href@noop {} {\bibfield  {journal} {\bibinfo  {journal} {ACS nano}\
  }\textbf {\bibinfo {volume} {8}},\ \bibinfo {pages} {4033} (\bibinfo {year}
  {2014})}\BibitemShut {NoStop}%
\bibitem [{\citenamefont {Buscema}\ \emph {et~al.}(2014)\citenamefont
  {Buscema}, \citenamefont {Groenendijk}, \citenamefont {Blanter},
  \citenamefont {Steele}, \citenamefont {Van Der~Zant},\ and\ \citenamefont
  {Castellanos-Gomez}}]{buscema2014fast}%
  \BibitemOpen
  \bibfield  {author} {\bibinfo {author} {\bibfnamefont {M.}~\bibnamefont
  {Buscema}}, \bibinfo {author} {\bibfnamefont {D.~J.}\ \bibnamefont
  {Groenendijk}}, \bibinfo {author} {\bibfnamefont {S.~I.}\ \bibnamefont
  {Blanter}}, \bibinfo {author} {\bibfnamefont {G.~A.}\ \bibnamefont {Steele}},
  \bibinfo {author} {\bibfnamefont {H.~S.}\ \bibnamefont {Van Der~Zant}}, \
  and\ \bibinfo {author} {\bibfnamefont {A.}~\bibnamefont
  {Castellanos-Gomez}},\ }\href@noop {} {\bibfield  {journal} {\bibinfo
  {journal} {Nano letters}\ }\textbf {\bibinfo {volume} {14}},\ \bibinfo
  {pages} {3347} (\bibinfo {year} {2014})}\BibitemShut {NoStop}%
\bibitem [{\citenamefont {Castellanos-Gomez}\ \emph {et~al.}(2014)\citenamefont
  {Castellanos-Gomez}, \citenamefont {Vicarelli}, \citenamefont {Prada},
  \citenamefont {Island}, \citenamefont {Narasimha-Acharya}, \citenamefont
  {Blanter}, \citenamefont {Groenendijk}, \citenamefont {Buscema},
  \citenamefont {Steele}, \citenamefont {Alvarez} \emph
  {et~al.}}]{castellanos2014isolation}%
  \BibitemOpen
  \bibfield  {author} {\bibinfo {author} {\bibfnamefont {A.}~\bibnamefont
  {Castellanos-Gomez}}, \bibinfo {author} {\bibfnamefont {L.}~\bibnamefont
  {Vicarelli}}, \bibinfo {author} {\bibfnamefont {E.}~\bibnamefont {Prada}},
  \bibinfo {author} {\bibfnamefont {J.~O.}\ \bibnamefont {Island}}, \bibinfo
  {author} {\bibfnamefont {K.}~\bibnamefont {Narasimha-Acharya}}, \bibinfo
  {author} {\bibfnamefont {S.~I.}\ \bibnamefont {Blanter}}, \bibinfo {author}
  {\bibfnamefont {D.~J.}\ \bibnamefont {Groenendijk}}, \bibinfo {author}
  {\bibfnamefont {M.}~\bibnamefont {Buscema}}, \bibinfo {author} {\bibfnamefont
  {G.~A.}\ \bibnamefont {Steele}}, \bibinfo {author} {\bibfnamefont
  {J.}~\bibnamefont {Alvarez}},  \emph {et~al.},\ }\href@noop {} {\bibfield
  {journal} {\bibinfo  {journal} {2D Materials}\ }\textbf {\bibinfo {volume}
  {1}},\ \bibinfo {pages} {025001} (\bibinfo {year} {2014})}\BibitemShut
  {NoStop}%
\bibitem [{\citenamefont {Xia}\ \emph {et~al.}(2014)\citenamefont {Xia},
  \citenamefont {Wang},\ and\ \citenamefont {Jia}}]{xia2014rediscovering}%
  \BibitemOpen
  \bibfield  {author} {\bibinfo {author} {\bibfnamefont {F.}~\bibnamefont
  {Xia}}, \bibinfo {author} {\bibfnamefont {H.}~\bibnamefont {Wang}}, \ and\
  \bibinfo {author} {\bibfnamefont {Y.}~\bibnamefont {Jia}},\ }\href@noop {}
  {\bibfield  {journal} {\bibinfo  {journal} {Nature communications}\ }\textbf
  {\bibinfo {volume} {5}},\ \bibinfo {pages} {4458} (\bibinfo {year}
  {2014})}\BibitemShut {NoStop}%
\bibitem [{\citenamefont {Novoselov}\ \emph {et~al.}(2004)\citenamefont
  {Novoselov}, \citenamefont {Geim}, \citenamefont {Morozov}, \citenamefont
  {Jiang}, \citenamefont {Zhang}, \citenamefont {Dubonos}, \citenamefont
  {Grigorieva},\ and\ \citenamefont {Firsov}}]{novoselov2004electric}%
  \BibitemOpen
  \bibfield  {author} {\bibinfo {author} {\bibfnamefont {K.~S.}\ \bibnamefont
  {Novoselov}}, \bibinfo {author} {\bibfnamefont {A.~K.}\ \bibnamefont {Geim}},
  \bibinfo {author} {\bibfnamefont {S.~V.}\ \bibnamefont {Morozov}}, \bibinfo
  {author} {\bibfnamefont {D.}~\bibnamefont {Jiang}}, \bibinfo {author}
  {\bibfnamefont {Y.}~\bibnamefont {Zhang}}, \bibinfo {author} {\bibfnamefont
  {S.~V.}\ \bibnamefont {Dubonos}}, \bibinfo {author} {\bibfnamefont {I.~V.}\
  \bibnamefont {Grigorieva}}, \ and\ \bibinfo {author} {\bibfnamefont {A.~A.}\
  \bibnamefont {Firsov}},\ }\href@noop {} {\bibfield  {journal} {\bibinfo
  {journal} {science}\ }\textbf {\bibinfo {volume} {306}},\ \bibinfo {pages}
  {666} (\bibinfo {year} {2004})}\BibitemShut {NoStop}%
\bibitem [{\citenamefont {Tombros}\ \emph {et~al.}(2007)\citenamefont
  {Tombros}, \citenamefont {Jozsa}, \citenamefont {Popinciuc}, \citenamefont
  {Jonkman},\ and\ \citenamefont {Van~Wees}}]{tombros2007electronic}%
  \BibitemOpen
  \bibfield  {author} {\bibinfo {author} {\bibfnamefont {N.}~\bibnamefont
  {Tombros}}, \bibinfo {author} {\bibfnamefont {C.}~\bibnamefont {Jozsa}},
  \bibinfo {author} {\bibfnamefont {M.}~\bibnamefont {Popinciuc}}, \bibinfo
  {author} {\bibfnamefont {H.~T.}\ \bibnamefont {Jonkman}}, \ and\ \bibinfo
  {author} {\bibfnamefont {B.~J.}\ \bibnamefont {Van~Wees}},\ }\href@noop {}
  {\bibfield  {journal} {\bibinfo  {journal} {Nature}\ }\textbf {\bibinfo
  {volume} {448}},\ \bibinfo {pages} {571} (\bibinfo {year}
  {2007})}\BibitemShut {NoStop}%
\bibitem [{\citenamefont {Avsar}\ \emph {et~al.}(2017)\citenamefont {Avsar},
  \citenamefont {Tan}, \citenamefont {Kurpas}, \citenamefont {Gmitra},
  \citenamefont {Watanabe}, \citenamefont {Taniguchi}, \citenamefont {Fabian},\
  and\ \citenamefont {{\"O}zyilmaz}}]{avsar2017gate}%
  \BibitemOpen
  \bibfield  {author} {\bibinfo {author} {\bibfnamefont {A.}~\bibnamefont
  {Avsar}}, \bibinfo {author} {\bibfnamefont {J.~Y.}\ \bibnamefont {Tan}},
  \bibinfo {author} {\bibfnamefont {M.}~\bibnamefont {Kurpas}}, \bibinfo
  {author} {\bibfnamefont {M.}~\bibnamefont {Gmitra}}, \bibinfo {author}
  {\bibfnamefont {K.}~\bibnamefont {Watanabe}}, \bibinfo {author}
  {\bibfnamefont {T.}~\bibnamefont {Taniguchi}}, \bibinfo {author}
  {\bibfnamefont {J.}~\bibnamefont {Fabian}}, \ and\ \bibinfo {author}
  {\bibfnamefont {B.}~\bibnamefont {{\"O}zyilmaz}},\ }\href@noop {} {\bibfield
  {journal} {\bibinfo  {journal} {Nature Physics}\ }\textbf {\bibinfo {volume}
  {13}},\ \bibinfo {pages} {nphys4141} (\bibinfo {year} {2017})}\BibitemShut
  {NoStop}%
\bibitem [{\citenamefont {Li}\ \emph {et~al.}(2017)\citenamefont {Li},
  \citenamefont {Kim}, \citenamefont {Jin}, \citenamefont {Ye}, \citenamefont
  {Qiu}, \citenamefont {Felipe}, \citenamefont {Shi}, \citenamefont {Chen},
  \citenamefont {Zhang}, \citenamefont {Yang} \emph {et~al.}}]{li2017direct}%
  \BibitemOpen
  \bibfield  {author} {\bibinfo {author} {\bibfnamefont {L.}~\bibnamefont
  {Li}}, \bibinfo {author} {\bibfnamefont {J.}~\bibnamefont {Kim}}, \bibinfo
  {author} {\bibfnamefont {C.}~\bibnamefont {Jin}}, \bibinfo {author}
  {\bibfnamefont {G.~J.}\ \bibnamefont {Ye}}, \bibinfo {author} {\bibfnamefont
  {D.~Y.}\ \bibnamefont {Qiu}}, \bibinfo {author} {\bibfnamefont
  {H.}~\bibnamefont {Felipe}}, \bibinfo {author} {\bibfnamefont
  {Z.}~\bibnamefont {Shi}}, \bibinfo {author} {\bibfnamefont {L.}~\bibnamefont
  {Chen}}, \bibinfo {author} {\bibfnamefont {Z.}~\bibnamefont {Zhang}},
  \bibinfo {author} {\bibfnamefont {F.}~\bibnamefont {Yang}},  \emph {et~al.},\
  }\href@noop {} {\bibfield  {journal} {\bibinfo  {journal} {Nature
  nanotechnology}\ }\textbf {\bibinfo {volume} {12}},\ \bibinfo {pages} {21}
  (\bibinfo {year} {2017})}\BibitemShut {NoStop}%
\bibitem [{\citenamefont {Popovi{\'c}}\ \emph {et~al.}(2015)\citenamefont
  {Popovi{\'c}}, \citenamefont {Kurdestany},\ and\ \citenamefont
  {Satpathy}}]{popovic2015electronic}%
  \BibitemOpen
  \bibfield  {author} {\bibinfo {author} {\bibfnamefont {Z.}~\bibnamefont
  {Popovi{\'c}}}, \bibinfo {author} {\bibfnamefont {J.~M.}\ \bibnamefont
  {Kurdestany}}, \ and\ \bibinfo {author} {\bibfnamefont {S.}~\bibnamefont
  {Satpathy}},\ }\href@noop {} {\bibfield  {journal} {\bibinfo  {journal}
  {Physical Review B}\ }\textbf {\bibinfo {volume} {92}},\ \bibinfo {pages}
  {035135} (\bibinfo {year} {2015})}\BibitemShut {NoStop}%
\bibitem [{\citenamefont {Kurpas}\ \emph {et~al.}(2016)\citenamefont {Kurpas},
  \citenamefont {Gmitra},\ and\ \citenamefont {Fabian}}]{kurpas2016spin}%
  \BibitemOpen
  \bibfield  {author} {\bibinfo {author} {\bibfnamefont {M.}~\bibnamefont
  {Kurpas}}, \bibinfo {author} {\bibfnamefont {M.}~\bibnamefont {Gmitra}}, \
  and\ \bibinfo {author} {\bibfnamefont {J.}~\bibnamefont {Fabian}},\
  }\href@noop {} {\bibfield  {journal} {\bibinfo  {journal} {Physical Review
  B}\ }\textbf {\bibinfo {volume} {94}},\ \bibinfo {pages} {155423} (\bibinfo
  {year} {2016})}\BibitemShut {NoStop}%
\bibitem [{\citenamefont {Li}\ and\ \citenamefont
  {Appelbaum}(2014)}]{li2014electrons}%
  \BibitemOpen
  \bibfield  {author} {\bibinfo {author} {\bibfnamefont {P.}~\bibnamefont
  {Li}}\ and\ \bibinfo {author} {\bibfnamefont {I.}~\bibnamefont {Appelbaum}},\
  }\href@noop {} {\bibfield  {journal} {\bibinfo  {journal} {Physical Review
  B}\ }\textbf {\bibinfo {volume} {90}},\ \bibinfo {pages} {115439} (\bibinfo
  {year} {2014})}\BibitemShut {NoStop}%
\bibitem [{\citenamefont {Dyakonov}\ and\ \citenamefont
  {Perel}(1972)}]{dyakonov1972spin}%
  \BibitemOpen
  \bibfield  {author} {\bibinfo {author} {\bibfnamefont {M.}~\bibnamefont
  {Dyakonov}}\ and\ \bibinfo {author} {\bibfnamefont {V.}~\bibnamefont
  {Perel}},\ }\href@noop {} {\bibfield  {journal} {\bibinfo  {journal} {Soviet
  Physics Solid State, USSR}\ }\textbf {\bibinfo {volume} {13}},\ \bibinfo
  {pages} {3023} (\bibinfo {year} {1972})}\BibitemShut {NoStop}%
\bibitem [{\citenamefont {Fabian}\ \emph {et~al.}(2007)\citenamefont {Fabian},
  \citenamefont {Matos-Abiague}, \citenamefont {Ertler}, \citenamefont
  {Stano},\ and\ \citenamefont {Zutic}}]{fabian2007semiconductor}%
  \BibitemOpen
  \bibfield  {author} {\bibinfo {author} {\bibfnamefont {J.}~\bibnamefont
  {Fabian}}, \bibinfo {author} {\bibfnamefont {A.}~\bibnamefont
  {Matos-Abiague}}, \bibinfo {author} {\bibfnamefont {C.}~\bibnamefont
  {Ertler}}, \bibinfo {author} {\bibfnamefont {P.}~\bibnamefont {Stano}}, \
  and\ \bibinfo {author} {\bibfnamefont {I.}~\bibnamefont {Zutic}},\
  }\href@noop {} {\bibfield  {journal} {\bibinfo  {journal} {Acta Physica
  Slovaca}\ }\textbf {\bibinfo {volume} {57}},\ \bibinfo {pages} {565}
  (\bibinfo {year} {2007})}\BibitemShut {NoStop}%
\bibitem [{\citenamefont {Averkiev}\ and\ \citenamefont
  {Golub}(1999)}]{averkiev1999giant}%
  \BibitemOpen
  \bibfield  {author} {\bibinfo {author} {\bibfnamefont {N.}~\bibnamefont
  {Averkiev}}\ and\ \bibinfo {author} {\bibfnamefont {L.}~\bibnamefont
  {Golub}},\ }\href@noop {} {\bibfield  {journal} {\bibinfo  {journal}
  {Physical Review B}\ }\textbf {\bibinfo {volume} {60}},\ \bibinfo {pages}
  {15582} (\bibinfo {year} {1999})}\BibitemShut {NoStop}%
\bibitem [{\citenamefont {Winkler}(2003)}]{winkler2003spin}%
  \BibitemOpen
  \bibfield  {author} {\bibinfo {author} {\bibfnamefont {R.}~\bibnamefont
  {Winkler}},\ }\href@noop {} {\emph {\bibinfo {title} {Spin-orbit coupling
  effects in two-dimensional electron and hole systems}}},\ Vol.\ \bibinfo
  {volume} {191}\ (\bibinfo  {publisher} {Springer Science \& Business Media},\
  \bibinfo {year} {2003})\BibitemShut {NoStop}%
\bibitem [{\citenamefont {Giannozzi}\ \emph {et~al.}(2017)\citenamefont
  {Giannozzi}, \citenamefont {Andreussi}, \citenamefont {Brumme}, \citenamefont
  {Bunau}, \citenamefont {Nardelli}, \citenamefont {Calandra}, \citenamefont
  {Car}, \citenamefont {Cavazzoni}, \citenamefont {Ceresoli}, \citenamefont
  {Cococcioni}, \citenamefont {Colonna}, \citenamefont {Carnimeo},
  \citenamefont {Corso}, \citenamefont {de~Gironcoli}, \citenamefont {Delugas},
  \citenamefont {Jr}, \citenamefont {Ferretti}, \citenamefont {Floris},
  \citenamefont {Fratesi}, \citenamefont {Fugallo}, \citenamefont {Gebauer},
  \citenamefont {Gerstmann}, \citenamefont {Giustino}, \citenamefont {Gorni},
  \citenamefont {Jia}, \citenamefont {Kawamura}, \citenamefont {Ko},
  \citenamefont {Kokalj}, \citenamefont {Küçükbenli}, \citenamefont
  {Lazzeri}, \citenamefont {Marsili}, \citenamefont {Marzari}, \citenamefont
  {Mauri}, \citenamefont {Nguyen}, \citenamefont {Nguyen}, \citenamefont {de-la
  Roza}, \citenamefont {Paulatto}, \citenamefont {Poncé}, \citenamefont
  {Rocca}, \citenamefont {Sabatini}, \citenamefont {Santra}, \citenamefont
  {Schlipf}, \citenamefont {Seitsonen}, \citenamefont {Smogunov}, \citenamefont
  {Timrov}, \citenamefont {Thonhauser}, \citenamefont {Umari}, \citenamefont
  {Vast}, \citenamefont {Wu},\ and\ \citenamefont {Baroni}}]{QE-2017}%
  \BibitemOpen
  \bibfield  {author} {\bibinfo {author} {\bibfnamefont {P.}~\bibnamefont
  {Giannozzi}}, \bibinfo {author} {\bibfnamefont {O.}~\bibnamefont
  {Andreussi}}, \bibinfo {author} {\bibfnamefont {T.}~\bibnamefont {Brumme}},
  \bibinfo {author} {\bibfnamefont {O.}~\bibnamefont {Bunau}}, \bibinfo
  {author} {\bibfnamefont {M.~B.}\ \bibnamefont {Nardelli}}, \bibinfo {author}
  {\bibfnamefont {M.}~\bibnamefont {Calandra}}, \bibinfo {author}
  {\bibfnamefont {R.}~\bibnamefont {Car}}, \bibinfo {author} {\bibfnamefont
  {C.}~\bibnamefont {Cavazzoni}}, \bibinfo {author} {\bibfnamefont
  {D.}~\bibnamefont {Ceresoli}}, \bibinfo {author} {\bibfnamefont
  {M.}~\bibnamefont {Cococcioni}}, \bibinfo {author} {\bibfnamefont
  {N.}~\bibnamefont {Colonna}}, \bibinfo {author} {\bibfnamefont
  {I.}~\bibnamefont {Carnimeo}}, \bibinfo {author} {\bibfnamefont {A.~D.}\
  \bibnamefont {Corso}}, \bibinfo {author} {\bibfnamefont {S.}~\bibnamefont
  {de~Gironcoli}}, \bibinfo {author} {\bibfnamefont {P.}~\bibnamefont
  {Delugas}}, \bibinfo {author} {\bibfnamefont {R.~A.~D.}\ \bibnamefont {Jr}},
  \bibinfo {author} {\bibfnamefont {A.}~\bibnamefont {Ferretti}}, \bibinfo
  {author} {\bibfnamefont {A.}~\bibnamefont {Floris}}, \bibinfo {author}
  {\bibfnamefont {G.}~\bibnamefont {Fratesi}}, \bibinfo {author} {\bibfnamefont
  {G.}~\bibnamefont {Fugallo}}, \bibinfo {author} {\bibfnamefont
  {R.}~\bibnamefont {Gebauer}}, \bibinfo {author} {\bibfnamefont
  {U.}~\bibnamefont {Gerstmann}}, \bibinfo {author} {\bibfnamefont
  {F.}~\bibnamefont {Giustino}}, \bibinfo {author} {\bibfnamefont
  {T.}~\bibnamefont {Gorni}}, \bibinfo {author} {\bibfnamefont
  {J.}~\bibnamefont {Jia}}, \bibinfo {author} {\bibfnamefont {M.}~\bibnamefont
  {Kawamura}}, \bibinfo {author} {\bibfnamefont {H.-Y.}\ \bibnamefont {Ko}},
  \bibinfo {author} {\bibfnamefont {A.}~\bibnamefont {Kokalj}}, \bibinfo
  {author} {\bibfnamefont {E.}~\bibnamefont {Küçükbenli}}, \bibinfo {author}
  {\bibfnamefont {M.}~\bibnamefont {Lazzeri}}, \bibinfo {author} {\bibfnamefont
  {M.}~\bibnamefont {Marsili}}, \bibinfo {author} {\bibfnamefont
  {N.}~\bibnamefont {Marzari}}, \bibinfo {author} {\bibfnamefont
  {F.}~\bibnamefont {Mauri}}, \bibinfo {author} {\bibfnamefont {N.~L.}\
  \bibnamefont {Nguyen}}, \bibinfo {author} {\bibfnamefont {H.-V.}\
  \bibnamefont {Nguyen}}, \bibinfo {author} {\bibfnamefont {A.~O.}\
  \bibnamefont {de-la Roza}}, \bibinfo {author} {\bibfnamefont
  {L.}~\bibnamefont {Paulatto}}, \bibinfo {author} {\bibfnamefont
  {S.}~\bibnamefont {Poncé}}, \bibinfo {author} {\bibfnamefont
  {D.}~\bibnamefont {Rocca}}, \bibinfo {author} {\bibfnamefont
  {R.}~\bibnamefont {Sabatini}}, \bibinfo {author} {\bibfnamefont
  {B.}~\bibnamefont {Santra}}, \bibinfo {author} {\bibfnamefont
  {M.}~\bibnamefont {Schlipf}}, \bibinfo {author} {\bibfnamefont {A.~P.}\
  \bibnamefont {Seitsonen}}, \bibinfo {author} {\bibfnamefont {A.}~\bibnamefont
  {Smogunov}}, \bibinfo {author} {\bibfnamefont {I.}~\bibnamefont {Timrov}},
  \bibinfo {author} {\bibfnamefont {T.}~\bibnamefont {Thonhauser}}, \bibinfo
  {author} {\bibfnamefont {P.}~\bibnamefont {Umari}}, \bibinfo {author}
  {\bibfnamefont {N.}~\bibnamefont {Vast}}, \bibinfo {author} {\bibfnamefont
  {X.}~\bibnamefont {Wu}}, \ and\ \bibinfo {author} {\bibfnamefont
  {S.}~\bibnamefont {Baroni}},\ }\href
  {http://stacks.iop.org/0953-8984/29/i=46/a=465901} {\bibfield  {journal}
  {\bibinfo  {journal} {Journal of Physics: Condensed Matter}\ }\textbf
  {\bibinfo {volume} {29}},\ \bibinfo {pages} {465901} (\bibinfo {year}
  {2017})}\BibitemShut {NoStop}%
\bibitem [{\citenamefont {Takao}\ \emph {et~al.}(1981)\citenamefont {Takao},
  \citenamefont {Asahina},\ and\ \citenamefont {Morita}}]{takao1981electronic}%
  \BibitemOpen
  \bibfield  {author} {\bibinfo {author} {\bibfnamefont {Y.}~\bibnamefont
  {Takao}}, \bibinfo {author} {\bibfnamefont {H.}~\bibnamefont {Asahina}}, \
  and\ \bibinfo {author} {\bibfnamefont {A.}~\bibnamefont {Morita}},\
  }\href@noop {} {\bibfield  {journal} {\bibinfo  {journal} {Journal of the
  Physical Society of Japan}\ }\textbf {\bibinfo {volume} {50}},\ \bibinfo
  {pages} {3362} (\bibinfo {year} {1981})}\BibitemShut {NoStop}%
\bibitem [{\citenamefont {Qiao}\ \emph {et~al.}(2014)\citenamefont {Qiao},
  \citenamefont {Kong}, \citenamefont {Hu}, \citenamefont {Yang},\ and\
  \citenamefont {Ji}}]{qiao2014high}%
  \BibitemOpen
  \bibfield  {author} {\bibinfo {author} {\bibfnamefont {J.}~\bibnamefont
  {Qiao}}, \bibinfo {author} {\bibfnamefont {X.}~\bibnamefont {Kong}}, \bibinfo
  {author} {\bibfnamefont {Z.-X.}\ \bibnamefont {Hu}}, \bibinfo {author}
  {\bibfnamefont {F.}~\bibnamefont {Yang}}, \ and\ \bibinfo {author}
  {\bibfnamefont {W.}~\bibnamefont {Ji}},\ }\href@noop {} {\bibfield  {journal}
  {\bibinfo  {journal} {Nature communications}\ }\textbf {\bibinfo {volume}
  {5}},\ \bibinfo {pages} {4475} (\bibinfo {year} {2014})}\BibitemShut
  {NoStop}%
\bibitem [{\citenamefont {Averkiev}\ \emph {et~al.}(2002)\citenamefont
  {Averkiev}, \citenamefont {Golub},\ and\ \citenamefont
  {Willander}}]{averkiev2002spin}%
  \BibitemOpen
  \bibfield  {author} {\bibinfo {author} {\bibfnamefont {N.}~\bibnamefont
  {Averkiev}}, \bibinfo {author} {\bibfnamefont {L.}~\bibnamefont {Golub}}, \
  and\ \bibinfo {author} {\bibfnamefont {M.}~\bibnamefont {Willander}},\
  }\href@noop {} {\bibfield  {journal} {\bibinfo  {journal} {Journal of
  physics: condensed matter}\ }\textbf {\bibinfo {volume} {14}},\ \bibinfo
  {pages} {R271} (\bibinfo {year} {2002})}\BibitemShut {NoStop}%
\bibitem [{\citenamefont {Ando}\ \emph {et~al.}(1982)\citenamefont {Ando},
  \citenamefont {Fowler},\ and\ \citenamefont {Stern}}]{ando1982electronic}%
  \BibitemOpen
  \bibfield  {author} {\bibinfo {author} {\bibfnamefont {T.}~\bibnamefont
  {Ando}}, \bibinfo {author} {\bibfnamefont {A.~B.}\ \bibnamefont {Fowler}}, \
  and\ \bibinfo {author} {\bibfnamefont {F.}~\bibnamefont {Stern}},\
  }\href@noop {} {\bibfield  {journal} {\bibinfo  {journal} {Reviews of Modern
  Physics}\ }\textbf {\bibinfo {volume} {54}},\ \bibinfo {pages} {437}
  (\bibinfo {year} {1982})}\BibitemShut {NoStop}%
\bibitem [{\citenamefont {Yuan}\ \emph {et~al.}(2015)\citenamefont {Yuan},
  \citenamefont {Rudenko},\ and\ \citenamefont
  {Katsnelson}}]{yuan2015transport}%
  \BibitemOpen
  \bibfield  {author} {\bibinfo {author} {\bibfnamefont {S.}~\bibnamefont
  {Yuan}}, \bibinfo {author} {\bibfnamefont {A.}~\bibnamefont {Rudenko}}, \
  and\ \bibinfo {author} {\bibfnamefont {M.}~\bibnamefont {Katsnelson}},\
  }\href@noop {} {\bibfield  {journal} {\bibinfo  {journal} {Physical Review
  B}\ }\textbf {\bibinfo {volume} {91}},\ \bibinfo {pages} {115436} (\bibinfo
  {year} {2015})}\BibitemShut {NoStop}%
\bibitem [{\citenamefont {Liu}\ \emph {et~al.}(2016)\citenamefont {Liu},
  \citenamefont {Low},\ and\ \citenamefont {Ruden}}]{liu2016mobility}%
  \BibitemOpen
  \bibfield  {author} {\bibinfo {author} {\bibfnamefont {Y.}~\bibnamefont
  {Liu}}, \bibinfo {author} {\bibfnamefont {T.}~\bibnamefont {Low}}, \ and\
  \bibinfo {author} {\bibfnamefont {P.~P.}\ \bibnamefont {Ruden}},\ }\href@noop
  {} {\bibfield  {journal} {\bibinfo  {journal} {Physical Review B}\ }\textbf
  {\bibinfo {volume} {93}},\ \bibinfo {pages} {165402} (\bibinfo {year}
  {2016})}\BibitemShut {NoStop}%
\bibitem [{\citenamefont {Wang}\ \emph {et~al.}(2015)\citenamefont {Wang},
  \citenamefont {Jones}, \citenamefont {Seyler}, \citenamefont {Tran},
  \citenamefont {Jia}, \citenamefont {Zhao}, \citenamefont {Wang},
  \citenamefont {Yang}, \citenamefont {Xu},\ and\ \citenamefont
  {Xia}}]{wang2015highly}%
  \BibitemOpen
  \bibfield  {author} {\bibinfo {author} {\bibfnamefont {X.}~\bibnamefont
  {Wang}}, \bibinfo {author} {\bibfnamefont {A.~M.}\ \bibnamefont {Jones}},
  \bibinfo {author} {\bibfnamefont {K.~L.}\ \bibnamefont {Seyler}}, \bibinfo
  {author} {\bibfnamefont {V.}~\bibnamefont {Tran}}, \bibinfo {author}
  {\bibfnamefont {Y.}~\bibnamefont {Jia}}, \bibinfo {author} {\bibfnamefont
  {H.}~\bibnamefont {Zhao}}, \bibinfo {author} {\bibfnamefont {H.}~\bibnamefont
  {Wang}}, \bibinfo {author} {\bibfnamefont {L.}~\bibnamefont {Yang}}, \bibinfo
  {author} {\bibfnamefont {X.}~\bibnamefont {Xu}}, \ and\ \bibinfo {author}
  {\bibfnamefont {F.}~\bibnamefont {Xia}},\ }\href@noop {} {\bibfield
  {journal} {\bibinfo  {journal} {Nature nanotechnology}\ }\textbf {\bibinfo
  {volume} {10}},\ \bibinfo {pages} {517} (\bibinfo {year} {2015})}\BibitemShut
  {NoStop}%
\bibitem [{\citenamefont {L{\"o}wdin}(1951)}]{lowdin1951note}%
  \BibitemOpen
  \bibfield  {author} {\bibinfo {author} {\bibfnamefont {P.-O.}\ \bibnamefont
  {L{\"o}wdin}},\ }\href@noop {} {\bibfield  {journal} {\bibinfo  {journal}
  {The Journal of Chemical Physics}\ }\textbf {\bibinfo {volume} {19}},\
  \bibinfo {pages} {1396} (\bibinfo {year} {1951})}\BibitemShut {NoStop}%
\end{thebibliography}%

\end{document}